\documentclass[prl,twocolumn,floatfix]{revtex4}
\usepackage{epsfig}
\begin{document}

\title{Optical control of Faraday rotation in hot Rb vapor}
\author{Paul Siddons}
\author{Charles S. Adams}
\author{Ifan G. Hughes}
\affiliation{Department of Physics,
Durham University, South Road, Durham, DH1~3LE, UK}

\date{\today}

\begin{abstract}
\noindent We demonstrate controlled polarization rotation of an optical field conditional on the presence of a second field. Induced rotations of greater than $\pi/2$~rad are seen with a transmission of $95\%$, corresponding to a ratio of phase shift to absorption of $40\pi$.  This combination of large, controlled rotation and low loss is well-suited for the manipulation of both classical and quantum light pulses.
    
\end{abstract}
\maketitle

The ability to manipulate optical pulses is central to the advancement of information and communications technology~\cite{nphoton}. All-optical switching~\cite{switchrev} has the advantage that the optical information can be processed without conversion to an electrical signal. An all-optical switch is produced by using an optical control field to modify the refractive index or the absorption of the medium, i.e., the real or the imaginary part of the electrical susceptibility, $\chi^{\mathrm R}$ and $\chi^{\mathrm I}$. For example, in electromagnetically induced transparency (EIT)~\cite{Harris98,YanPRA01} or off-resonance Raman resonances~\cite{ScullyPRL91,ScullyPRL92,YavuzPRL08} a strong control field is employed to reduce the absorption at a particular frequency.
Reducing the intensity of the control field to the single photon level is of interest for certain quantum information protocols~\cite{qi}. All-optical switching at low light levels  has been demonstrated using  EIT \cite{Zhang07}, and also using transverse optical pattern formation~\cite{GauthierSci,GauthierPRA08}.

An additional important criteria in an optical switch is high fidelity transmission of the input field to reduce the loss of information. The requirement of a large modulation depth concomitant with low absorption suggests that controlling the phase, or $\chi^{\mathrm R}$, is preferable, as in electro-optic devices such as the Mach-Zehnder modulator~\cite{MZ}. In this case, the figure of merit of the switching process is characterized by the change in the birefringence of the medium divided by absorption, $\Delta\chi^{\mathrm R}/\chi^{\mathrm I}$, i.e., the ratio of the phase shift to the optical depth, $2\Delta\phi/\mathrm{OD}$. We note that for a two-state resonance, the Kramers-Kronig relations  show that this ratio is largest far from resonance where the dispersion is also smaller~\cite{Siddons09}. The change in $\chi^{\mathrm R}$ can be between different polarization modes of the light field giving rise to birefringence or Faraday rotation. Polarization rotation of a linearly polarized optical field has been studied extensively in atomic systems. Such rotations may be induced by an applied magnetic field, i.e., the Faraday effect~\cite{Bud, F91, SiddonsNature}, by an applied electric field~\cite{fstark}, or by spin polarizing the medium, i.e., the paramagnetic Faraday effect~\cite{Happer67,Wieman76,Buell,Hammerer,polzik}. For optical switching, a rotation angle of $\pi/2$~rad is required such that two orthogonal linear polarization modes can be exchanged.  An EIT scheme reported by Li \textit{et al.}\ provides rotations in the region of $\pi/4$~rad, with $\sim50\%$ loss~\cite{LiPRA06}. Larger rotations at lower loss where seen by Siddons \textit{et al.}\ using the off-resonant Faraday effect~\cite{SiddonsNature}, but without optical control.

In this letter we demonstrate that high fidelity modulation ($>90\%$) of the input field using optical control in an atomic vapor. To achieve a large induced rotation with low loss, i.e., a high value of the figure of merit, $\Delta\chi^{\mathrm R}/\chi^{\mathrm I}$, we bias the rotation of the probe using the off-resonant Faraday effect and employ a control beam to induce population transfer to modulate around this bias. We demonstrate optical control of the Faraday rotation due to both changes in the total number of atoms and due to their spin distribution. For the probe field detuned by more than 5 times the inhomogeneous atomic linewidth we observe a phase shift of $\pi/2$ radians with a loss of less than $5\%$, corresponding to $\Delta\chi^{\mathrm R}/\chi^{\mathrm I}=40\pi$. This combination of large dispersion and low loss is interesting in the context of all optical manipulation of both classical and quantum light pulses, for example, all-optical single qubit 
rotations for photons~\cite{Petro05}. As a large rotation is achieved off-resonance the process potentially can be operated at high bandwidth of order GHz~\cite{SiddonsNature}. In addition, by combining this technique with the dispersive filtering properties of the Faraday effect \cite{F91,F93,Abel09} one could realize an optically tunable narrowband filter.

\begin{figure}[tbh]
\centering
\includegraphics*[width=8.0cm]{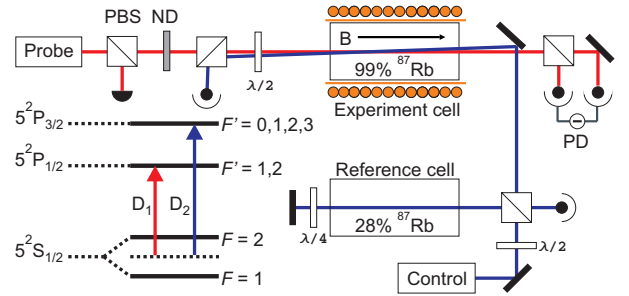}
\caption{(Color online). Schematic of the experimental apparatus.  A probe beam passes through a polarization beam splitter (PBS), providing linearly polarized light.  The beam is attenuated with a neutral density filter (ND) before passing through a heated vapor cell. A half-wave plate ($\lambda/2$) is used to control the polarization angle of the light before it is analyzed with a PBS and collected on a differencing photo-diode (PD).  A control beam is linearly polarized and counter-propagated at a small crossing angle.  A small fraction of the beam is used to perform sub-Doppler spectroscopy in a reference cell.}
\label{fig:setup}
\end{figure}   

Figure~\ref{fig:setup} shows a schematic of the experimental apparatus along with the energy level scheme used to observe the optically controlled Faraday effect on the D$_1$ ($\mathrm{5^2S_{1/2}\rightarrow5^2P_{1/2}}$) transition of rubidium.  The source of probe light was an external cavity diode laser (ECDL) at 795~nm.  The probe laser output polarization was linearly polarized and attenuated to be less than $1\:\mu$W.  The beam had a $1/\mathrm{e}^2$ radius of 0.8~mm.  After passing through a half-wave plate the beam was sent through a 75~mm heated vapor cell containing the Rb isotopes according to the ratio $^{87}$Rb:$^{85}$Rb of 99:1.  Heating and magnetic field was provided by a solenoid, based on the design of Ref.~\cite{McCarron07}.  Upon transmission through the cell, the two orthogonal linear polarizations of the beam were separated with a polarizing beam splitter (PBS) cube and sent to a differencing photo-diode.  For balanced detection of rotation the polarization plane was set to an angle of $\pi/4$ to the axis of the analyzing PBS cube~\cite{Huard}.  

\begin{figure}[tbh]
\centering
\includegraphics*[width=8.0cm]{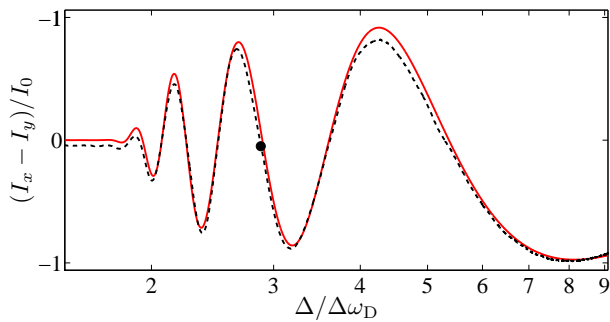}
\caption{(Color online).  Probe differencing signal produced by scanning the probe versus red detuning, $\Delta$, from the D$_1$ $^{87}$Rb $F=2\rightarrow F^\prime=1$ transition in units of Doppler width $\Delta\omega_{\mathrm{D}}=2\pi\times 571\:\mathrm{MHz}$.  The dashed black curve shows measured data, whilst the red curve shows the signal expected from theory.  The temperature of the cell is $115^\circ$C and the applied magnetic field is 204~G. }
\label{fig:fixedD1}
\end{figure}

Figure~\ref{fig:fixedD1} shows the effect of an applied magnetic field on the polarization of the probe beam transmitted through the cell.  The signal detected is the intensity difference of the two orthogonal linear polarizations $I_x$ and $I_y$ normalized by the off-resonant intensity $I_0$, and is related to the polarization angle of the light~\cite{SiddonsNature}.  Oscillations in the signal are seen due to the spectral dependence of the Faraday rotation.  Figure~\ref{fig:fixedD1} also shows the calculated theoretical signal obtained by diagonalizing the complete Hamiltonian of the system. Good agreement with experimental data is seen: any difference is due to unbalanced photo detectors.

To investigate optical control of the Faraday rotation we fix the detuning of the probe laser at detuning where the polarimeter signal is close to zero (indicated by the dot in Fig.~\ref{fig:fixedD1}) and add an optical control field resonant with the D$_2$ line ($\mathrm{5^2S_{1/2}\rightarrow5^2P_{3/2}}$) at 780~nm.  The control beam was linearly polarized with a spot size of 2~mm ($1/\mathrm{e}^2$ radius), and counter-propagated through the cell at an angle of $\sim5$~mrad.  A natural abundance Rb cell was used as a frequency reference to calibrate the detuning of the control field relative to the D$_2$ line (see Fig.~~\ref{fig:setup}).

\begin{figure}[tbh]
\centering
\includegraphics*[width=8.0cm]{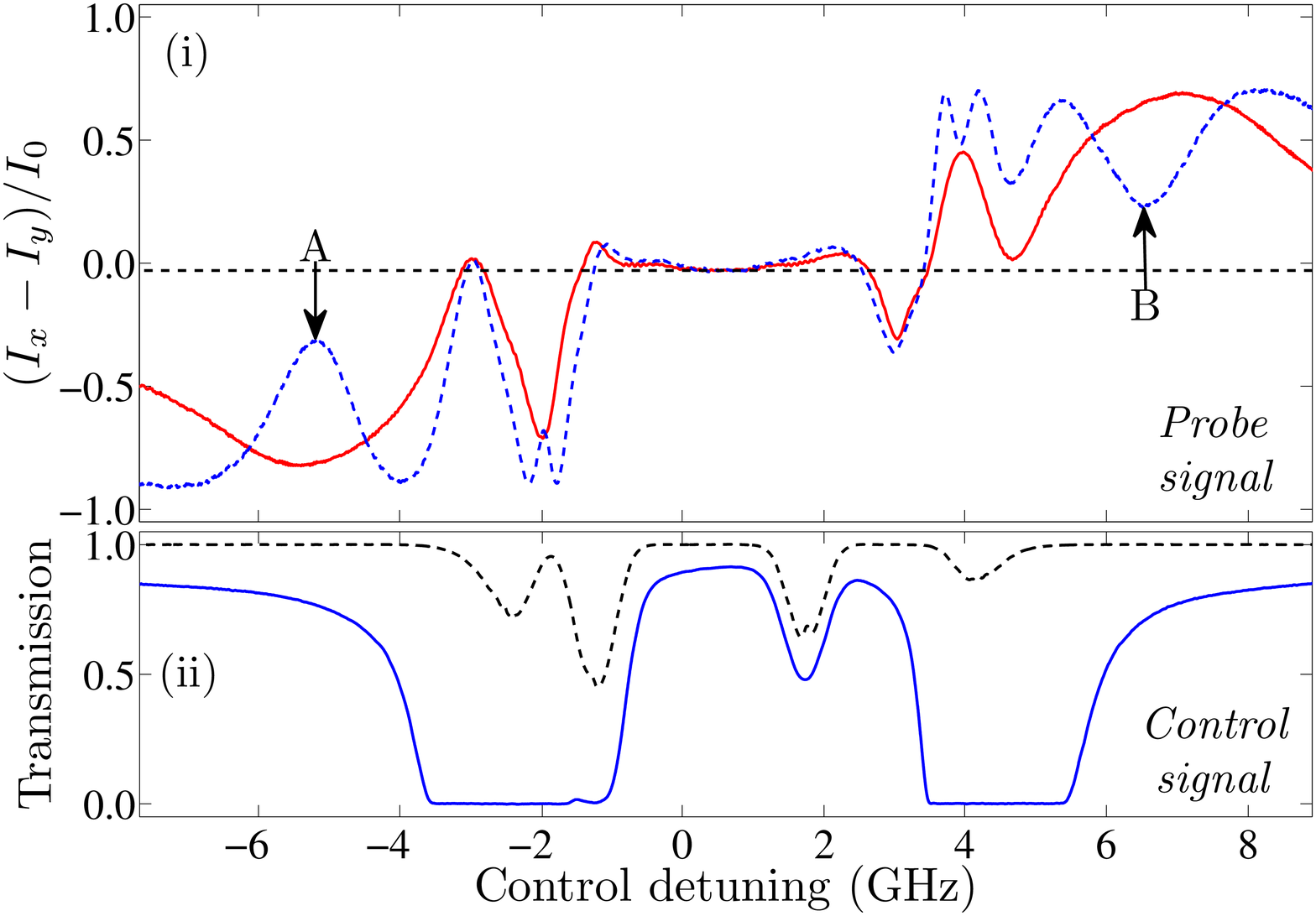}
\caption{(Color online). (i) Differencing signal of the probe with a control field of 9~mW (solid red) and 30~mW (dashed blue).  The experiment cell is at a temperature of $\sim115^\circ$C, with a 204~G applied magnetic field.  The probe is at a red detuning of $\sim2.9$ Doppler widths (1.7~GHz), marked in Fig.~\ref{fig:fixedD1}.  In the absence of the control field, the differencing signal is given by the dashed black line. (ii) Transmission of the control laser through the room temperature natural abundance reference cell (dashed black) and the $^{87}$Rb experiment cell (blue).     Control detuning is with respect to the weighted  D$_2$ line-center.}
\label{fig:power}
\end{figure}

In Figure~\ref{fig:power} we show the response of the Faraday rotation signal as a function of the frequency of the control field.  Figure~\ref{fig:power}(i) shows the differencing signal for the probe for the same temperature and magnetic field as Fig.~\ref{fig:fixedD1}.   Figure~\ref{fig:power}(ii) shows the transmission of the 2~mW control beam through the experiment cell and a weaker beam through the reference cell.  Between the two $^{87}$Rb absorption lines the control field appears to have little effect on the difference signal, but close to resonance and at greater detunings the effect of optical control is significant.   The maximum/minimum signal corresponds to alignment with the $x/y$ axis before folding back upon itself for greater rotation angles.  Increasing the control power increases the rotation angle whilst retaining similar spectral dependence, hence the dips seen in the 30~mW curve in Fig.~\ref{fig:power}(i) correspond to rotations beyond $\pi/4$~rad to the input beam, most noticeable at points A and B.  Note the magnitude of the maximum and minimum signals are not equal because the detectors in the differencing photo-diode are not perfectly balanced.

\begin{figure}[tbh]
\centering
\includegraphics*[width=8.0cm]{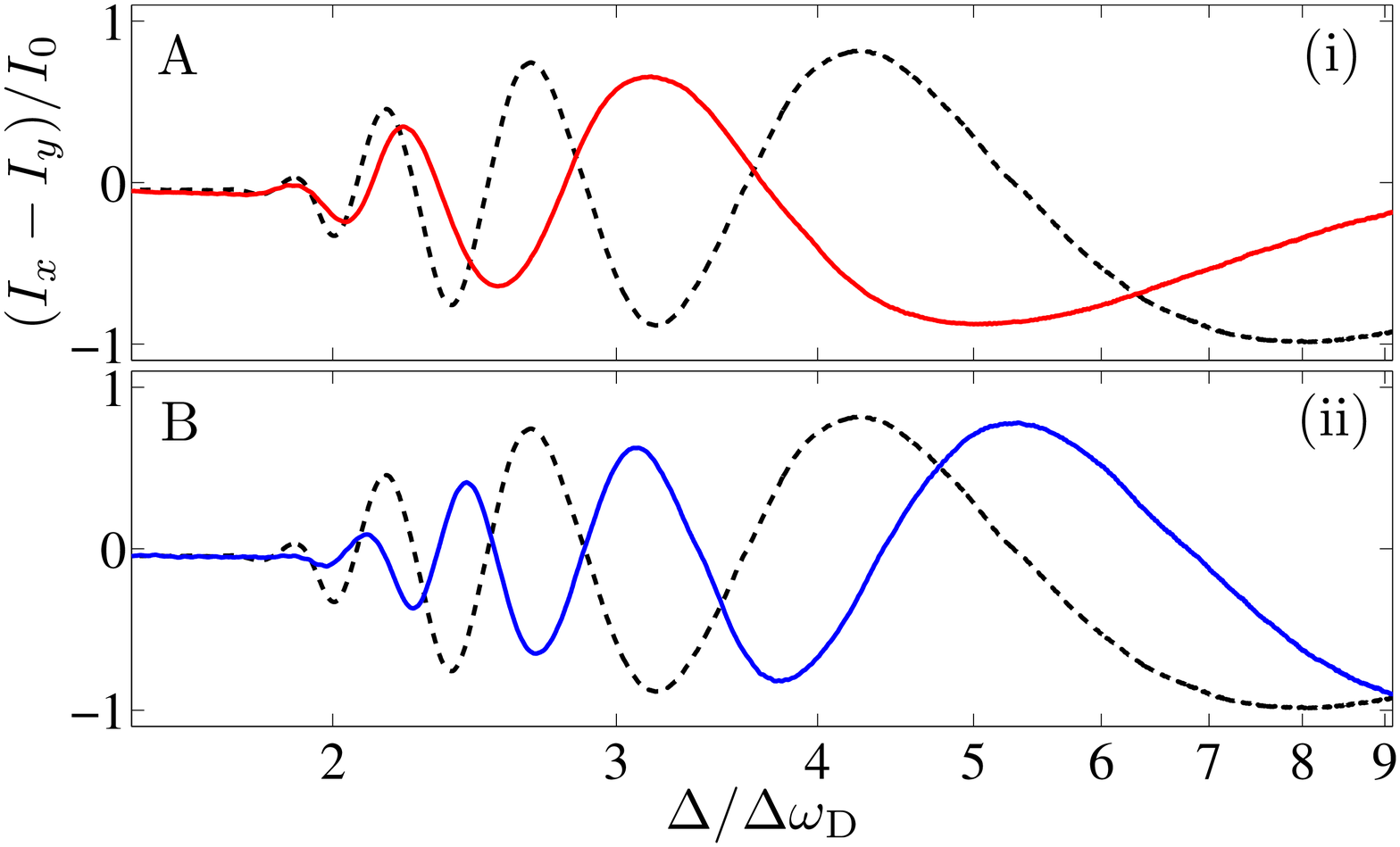}
\caption{(Color online). Probe differencing signal.   The dashed black curve shows the experimentally measured signal in the absence of the control field (From Figure~\ref{fig:fixedD1}).  Plots (i) and (ii) illustrate the effect of optical pumping on the probe signal when the 38~mW control field is fixed at detunings A and B given in Fig.~\ref{fig:power}.  Temperature and applied magnetic field are the same as Fig.~\ref{fig:fixedD1}.}
\label{fig:nemo}
\end{figure}

\begin{figure}[tbh]
\centering
\includegraphics*[width=8.0cm]{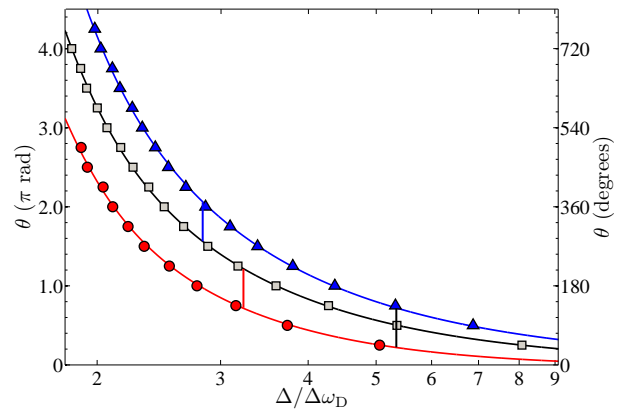}
\caption{(Color online). The measured rotation angles, $\theta$, from Fig.~\ref{fig:nemo} for no optical control (squares), a red-detuned control field (circles), and a blue-detuned control field (triangles).  The curves are from theory. Vertical bars show the detunings at which $\pi/2$ rotations are possible by switching amongst the three curves.}
\label{fig:theta}
\end{figure}

Extracting a rotation angle from this data is not trivial since the extrema of the signal depend on the transmission of the probe, which is spectrally dependent.  Instead, we fix the frequency of the control laser  and scan the probe light in a region red detuned from the D$_1$ line.  Figure~\ref{fig:nemo} shows the resulting difference signal produced in the presence of the control field.    Plots (i) and (ii) show the influence of the applied control field with its frequency fixed at the two points of maximum rotation shown in Fig.~\ref{fig:power}.  From these plots we are able to take the absolute angular rotation, $\theta$, of the probe using the zero-crossings and extrema (which are independent of transmission: see Ref.~\cite{SiddonsNature}).  The measured rotations are shown in Fig.~\ref{fig:theta}.  It can be seen that rotations of many $\pi$ radians are possible with the Faraday effect, as observed in previous studies \cite{SiddonsNature,Siddons09}.  For the rotation angle of $\pi/2$~rad induced by the applied magnetic and optical fields, the change in refractive index $\Delta n=5\times10^{-6}$, though changes of $10^{-4}$ and higher are possible for larger fields (see Ref.~\cite{SiddonsNature}).

In order to calculate the optically induced rotations we extended our steady-state model used to generate the theory curve in Fig.~\ref{fig:nemo} by setting the populations of the atomic states as independent parameters.  This model quantitatively imitates the behavior of the optical pumping inducing the controlled Faraday rotation.
Population transfer via a $\pi$ polarized pump is modeled as a change in the $F$ state populations; transfer by $\sigma^{\pm}$ pumping is modeled as an anisotropy in the $m_F$ state populations, the paramagnetic Faraday effect.  For the case of no pumping, an equilibrium population produces an excellent fit to data.  Decreasing the population of the $^{87}$Rb $F=2$ by 2.5\%, with an $m_F$ anisotropy such that there is an increased occupation of the $m_F=-2,-1$ states, reproduces the effect of a red-detuned control field; increasing the population by 16\%, with an $m_F$ anisotropy which increases the occupation of the $m_F=2,1$ states, reproduces the blue-detuned control field effect.  The parameters used here on an \textit{ad hoc} basis agree with the expected pumping behavior:  the red-detuned beam pumps depletes the population of the ground state being probed.  As such the rotation is decreased with respect to the equilibrium case.  The opposite is true for the blue-detuned beam.  The $m_F$ anisotropy is due to the pump polarization changing from its initially linear state to being highly elliptical as it propagates through the medium. 

\begin{figure}[tbh]
\centering
\includegraphics*[width=8.0cm]{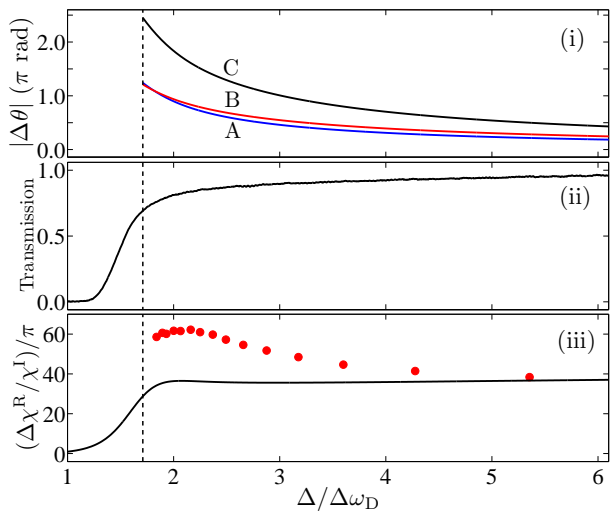}
\caption{(Color online).(i) The rotation difference between the cases with and without the optical control field, obtained from the experimental data shown in Fig.~\ref{fig:theta}.  Curves A (blue) and B (red) are, respectively, for blue and red detunings.  Curve C (black) shows the rotation difference between blue and red detunings.  (ii) The measured transmission of the linear probe in zero magnetic field.   (iii) The figure of merit of the switching scheme: the curve was obtained from theory, the points show measured data.  The dashed vertical line marks the limit close to resonance where the probe polarization becomes  elliptical due to circular dichroism (differential absorption of the left and right circularly polarized field components).}
\label{fig:deltatheta}
\end{figure}

By taking the difference of the curves in Fig.~\ref{fig:theta} we can obtain the rotation caused by switching amongst the three cases of no control field, and red/blue detuned control.  Curves A and B in Fig.~\ref{fig:deltatheta}(i) show the magnitude of the rotation between control field on and off, with $\pi/2$ rotation and $\sim90\%$ transmission for red detuning.  Note that the red- and blue-detuned cases have opposite sign, so that the difference between these two (curve C) has a greater magnitude, achieving $\pi/2$ at $\sim95\%$ transmission.  The transmission from Fig.~\ref{fig:deltatheta}(ii) was used to calculate the optical density, which in turn was used to calculate the figure of merit, shown in Fig.~\ref{fig:deltatheta}(iii).  The figure of merit is $>40\pi$ for detunings up to $5\Delta\omega_{\mathrm{D}}$.  This is more than an order of magnitude larger than previous work e.g.\ $1.4\pi$ for Ref.~\cite{LiPRA06}, and $\pi\times10^{-2}$ to $\pi\times10^{-1}$  for other experiments~\cite{Ent02,Choi07,YangJPB08}.  From the theory curve, the figure of merit is essentially constant beyond two Doppler widths from resonance.  This is ideal for broadband light where large differential dispersion over the spectrum of the pulse can lead to distortion~\cite{SiddonsNature}.  The theoretical and measured figure of merit do not agree close to zero detuning, since the probe beam was sufficiently strong to optically pump the medium, and alter its transmission through the medium; whereas the theory curve is based on the weak probe limit which in the case of the Rb D$_1$ line is 10~nW~\cite{Siddons08}.   

In summary, we have demonstrated the controlled polarization rotation of one optical field due to the presence of another, with high transmission of both beams.  A continuous-wave field was used to incoherently pump atoms into a dark ground state, a process which typically takes $0.1-1$~$\mu$s~\cite{Pearman02}.  Hence this process allows rapid switching, and has applications as a dynamic half-wave plate, a tunable Faraday dichroic beam splitter~\cite{Abel09}, or polarization modulation of single photons in a similar manner to the amplitude modulation performed using an EOM~\cite{Harris08}. 

Optical control of the Faraday effect could be used for all-optical single qubit 
rotations for photons~\cite{Petro05} and consequently opens new perspectives for 
all-optical quantum information processing. In the current experiment, a relatively strong control field is required.  In future work, a pulsed field will be used to coherently drive population into the excited state in a time less than the excited state lifetime.  From a simulation of population dynamics, only a small amount of anisotropy in the occupation of atomic states is required to observe rotations necessary to realize orthogonally polarized photon channels.  The nanosecond switching time, combined with the Gigahertz bandwidth off-resonant Faraday effect~\cite{SiddonsNature} could permit rapid high-fidelity switching at low light levels.

This work is supported by EPSRC.  We thank J Millen and M P A Jones for discussion, technical assistance and the loan of equipment.

\end{document}